\documentclass[12pt,aps,amsmath,amssymb,floatfix,nofootinbib]{revtex4}
\usepackage{amssymb,amsmath}
\newcommand{\CM}{{\mathbb C}}

\newcommand{\NM}{{\mathbb N}}

\newcommand{\ZM}{{\mathbb Z}}

\newcommand{\rn}{\overrightarrow{r}^{(n)}}
\newtheorem{theo}{Theorem}

\newtheorem{proposi}{Proposition}
\newtheorem{lemma}{Lemma}

\begin{document}
\title{On the Spectrum of the Resonant Quantum Kicked Rotor. }
\author{Italo Guarneri}
\address{
{\small  Center for Nonlinear and Complex Systems}\\
{\small  Universit\'a dell'Insubria, via Valleggio 11, I-22100 Como, Italy.}\\
{\small  Istituto Nazionale di Fisica Nucleare, Sezione di Pavia,
via Bassi 6, I-27100 Pavia, Italy.} }


\date{\today}


\begin{abstract}
 It is proven that none of the bands in the quasi-energy spectrum of the Quantum Kicked Rotor is flat at any primitive resonance of any order. Perturbative estimates of bandwidths at small kick strength are established  for the case of primitive resonances of prime order. Different bands scale with different powers of the kicking strength, due to degeneracies in the spectrum of the free rotor.
\end{abstract}


\maketitle

\subsection{Introduction.}

\subsubsection{Background.}

\noindent   The Quantum Kicked Rotor (QKR) is a famous model in the study
of quantum mechanical implications of classical chaotic dynamics \cite{CCIF79, Fish,Iz90}
. Its discrete-time dynamics is generated
in $L^2([0,2\pi])$ by the unitary propagator
\begin{equation}
\label{kr} {\hat
U}_{\beta,\tau,\mu}\;=\;e^{-i\frac{\tau}2(-i\frac d{d\theta}
+\beta)^2}\;e^{-i\mu\cos(\theta)}\;,
\end{equation}
where $\mu$ , $\beta$, and  $\tau$ are real parameters,
 and periodic boundary conditions are understood for the momentum operator
$-id/d\theta$.
The QKR model owes much of its fame to
dynamical localization,
which is expected to occur whenever $\tau$ is sufficiently incommensurate to
$2\pi$. Prior to lately established exact results \cite{BG02, Ji02}, this expectation has rested a long time  on  vast numerical evidence, and on formal assimilation of the QKR to a tight-binding model
of motion in a disordered potential \cite{Fish, FGP82}.
The so-called QKR resonances \cite{IS80,CG84,Fish,WGF03,CS86,DD06} occur in the opposite case, when $\tau$ is commensurate to $2\pi$, and appear to pose a simpler problem because the QKR dynamics is then
assimilated to motion in a periodic potential. A band spectrum is thus expected
of the propagator in (\ref{kr}),
hence a non-empty absolutely continuous spectrum, provided one at least of the bands
is not flat (\emph{i.e.}, reduced to a single, infinitely degenerate eigenvalue).  Precise definitions are given in Section \ref{stres}. Although band spectra at QKR resonances, and dynamical consequences thereof, were quite early discovered
\cite{IS80}, very little is  known  about the band structure, which is largely determined by the arithmetics of Gaussian sums. In particular, though no instance is known of "flat bands" besides the so-called "anti-resonance" that is observed at order $1$, their possible occurrence at higher orders has never been ruled
out. Moreover, analysis has been mostly restricted
to the "traditional"
 KR model, which has $\beta=0$. It was pointed out in
\cite{DD06} that the generalized KR model (\ref{kr}) presents a much richer family  of resonances, and
a distinction naturally arises between "primitive"  and "non-primitive" resonances, which are defined in Sect.\ref{stres} below.\\
Absence of flat bands is proven in this paper for primitive resonances of all orders.\\
Model (\ref{kr}) may be generalized by replacing $\cos(\theta)$ by some other "kicking potential"  $V(\theta)$. Resonances for such models have been studied in ref.\cite{DD06}. In ref. \cite{CG84} it was noted that flat bands at resonances are non-generic (in the sense of Baire) if $V(\theta)$ is chosen in the class of analytic potentials. Since  in such models pure AC spectrum for a dense set of $\tau$ entails some continuous spectral component (presumably singular) whenever $\tau/2\pi$ is sufficiently rapidly approximated by rationals, the conclusion followed, that the latter property is generic in that class of models. The same property is now proven to hold for the straight KR model (\ref{kr}) itself  as a byproduct of the present result.\\
Current knowledge of parametric dependence of bandwidths is mostly
about the asymptotic regime $1<<\mu<<q$, where bands are expected to be exponentially
small. This  behavior consistently matches the exponential localization of eigenfunctions in momentum space \cite{Fish,BG02},
which is known to occur in the QKR model whenever  $\tau/(2\pi)$
is sufficiently irrational (in the sense of rational approximation),
and is heuristically understood \cite{Iz93} on the grounds of a similarity of the
QKR to tight-binding models of Solid State physics\cite{Fish}.
In such models, following  Thouless
\cite{Thou}, bandwidth is related to conductance, so its exponential decay follows from Anderson localization.
\\
Little is known about parametric dependence in other parameter ranges. The case of not too large order is in particular relevant for the investigation of the nearly-resonant dynamics, in a variant of the QKR model where peculiar transport  phenomena are observed\cite{FGR,GR09}.
For small $\mu$, the dynamics (\ref{kr}) is a perturbation of a free rotation. However, perturbative analysis of the band structure has nontrivial aspects, because the unperturbed  quasi-energy spectrum is always degenerate at resonance, and  the degree of degeneracy depends on the arithmetics of the coprime integers $p,q$. The simplest case occurs when the divisor $q$ is prime, and is analyzed in this paper.

\subsubsection{Statement of Results.}
\label{stres}
\noindent
A KR resonance is said to occur whenever ${\hat U}_{\beta,\tau,\mu}$  commutes with a
momentum translation ${\hat T}^{Q}$  ($Q$ a strictly positive integer), where
${\hat T}\psi(\theta)=e^{i\theta}\psi(\theta)$. The least such $Q$ may be termed the "length" of the resonance.  Necessary and sufficient conditions for KR resonances are easily established:
\begin{proposi}
\label{dado}
{{\rm (Dana, Dorofeev \cite{DD06})} \sl  ${\hat U}_{\beta,\tau,\mu}$ commutes with ${\hat T}^{Q}$  if, and only if,\\
  (i) $\tau = 2\pi \tfrac PQ$ with $P$ integer,\\
 (ii) $\beta = \tfrac{\nu}P + \tfrac Q2$ mod$(1)$,
with $\nu$ an arbitrary integer.}
\end{proposi}
\noindent The integer $Q$ will always be equal to the resonance length in this paper. The integer $q$ such that  $P/Q=p/q$ with $p,q$ coprime integers will be  termed the order of the resonance.  A resonance will be termed "primitive" if its length is equal to its order, $Q=q$.
At resonance, the Bloch index
$\vartheta:=\theta$ mod$(2\pi/Q)$ is conserved.
A unitary map $\mathfrak b$ from
$L^2([0,2\pi])$ onto $L^2([0,2\pi/Q])\otimes\CM^q$ is defined via:
$$
({\mathfrak b}{\psi})_j(\vartheta) \;=\; \psi(\vartheta+2\pi\tfrac{j-1}Q)\;\;,\;1\leq j\leq Q\;\;,\vartheta\in [0,2\pi/Q]\;.
$$
Straightforward calculation shows that ${\mathfrak b}{\hat U}{\mathfrak b}^{-1}$ has a direct integral
decomposition \cite{RS,IS80,CG84}:
$$
{\mathfrak b}{\hat U}{\mathfrak b}^{-1}\;=\;\int^\oplus_{[0,2\pi/Q]}d\vartheta\;{{\bf S}}(\vartheta)\;,
$$
that is,
$({\mathfrak b}{\hat
U}\psi)(\vartheta)= {{\bf S}}(\vartheta)({\mathfrak
b}\psi)(\vartheta)$, where ${{\bf S}}(\vartheta)$ is the unitary $Q\times Q$ matrix
, which has matrix elements :
\begin{gather}
\label{def1}
{{S}}_{jk}(\vartheta)\;=\;e^{-i\mu\cos(\vartheta+2\pi(j-1)/Q)}\;{G}_{jk}\;,\\
\label{def2}
{G}_{jk}\;=\;\tfrac 1Q\sum\limits_{s=0}^{Q-1}e^{2\pi i s(j-k)/Q}\;a_{s+1}\;,\\
\label{def3}
a_r\;=\;e^{-i\pi p(r+\beta-1)^2/q}\;\;,\;\;1\leq r\leq Q\
\end{gather}
Of all parameters $p,q,\beta,\mu, \vartheta$, only those which are strictly necessary at a given time will be specified when listing the  arguments on which a quantity depends. The
numbers $a_r$ in (\ref{def3}) are the eigenvalues of the matrix ${\bf G}={\bf S}(\vartheta,0)$, which is defined  in (\ref{def2}). The eigenvalues of ${{\bf S}}(\vartheta,\mu)$ will be denoted $w_j=w_j(\vartheta,\mu)$ and numbered so that $ w_j(\vartheta, 0)=a_j$, ($j=1,2,\ldots,Q$). As $\vartheta$ varies in $(0,2\pi/Q)$, each eigenangle $\arg(w_j(\vartheta))$ sweeps a band in the quasi-energy spectrum of $\hat U$, that will be said to be "flat" in the case of a constant eigenangle. \\
Analysis of the band structure is easily performed in the case of primitive resonances with $Q=q=1$ , as they yield
one  band : $w_1(\vartheta,\mu)=e^{-i\pi p\beta^2}e^{-i\mu\cos(\vartheta)}$, which is not flat (if $\mu\neq 0$). The non-primitive resonance of order $q=1$ with  $p=1$, $Q=2$, $\beta=0$ is quite easily seen to yield flat bands and is known long since \cite{IS80} under the denotation of "anti-resonance". Flatness of all bands has been proven for all non-primitive resonances of order $1$ in ref.\cite{DD06} and can be verified by direct calculation of the matrix (\ref{def1}).   \\
Here it will be proven that:
\begin{theo}
\label{bds}
{No band is flat at any  primitive resonance of any order, as long as  $\mu\neq 0$. Hence, the spectrum of ${\hat
U}_{\beta,\tau,\mu}$, $(\mu\neq 0)$, is purely absolutely continuous at primitive resonances.
}
\end{theo}
Perturbative estimates of bandwidths will be proven for small $\mu$ in a special class of resonances . Perturbation theory is cumbersome, because the unperturbed ($\mu=0$) spectrum  is always degenerate, to an extent that depends on the arithmetics of $p$ and $q$. The simplest case is that of primitive resonances of prime order:
\begin{theo}
\label{bdwth}
{If $\tau=2\pi\tfrac pq$,  $q>2$ is prime,  $p$ is not a multiple of $q$, and $\beta=\tfrac 12$, then $a_r\;=\;a_{q-r+1}$ holds for $1\leq r\leq q$ and  all eigenvalues (\ref{def3}), except one, are twice degenerate. Asymptotically as $\mu\to 0$:
\begin{gather}
\frac{dw_j(\vartheta,\mu)}{d\vartheta}\;\sim\;\mu^{\alpha_j}s_j(p,q)\;
\sin(\vartheta)\;,\nonumber\\
\tfrac q2\;<\;\alpha_j=\mbox{\rm max}\{2j-1,q-2j+1\}\;\leq\;q\;.
\label{claim0}
\end{gather}
There are
positive constants $C, \gamma$ so that, asymptotically as $q\to\infty$ :
\begin{equation}
\label{claim3}
|s_j(p,q)|\;\lesssim\;Ce^{-\gamma q+A_j(q)}\;\;,\;\;\;\;\;A_j(q)=
O\left(\sqrt{q\log^3(q)}\right)\;.
\end{equation}
}
\end{theo}
The present proof yields an estimate $\gamma\geq 0.0016$, which  is  hardly optimal. Numerical calculations, though of necessity
restricted to small values of $q$, suggest an average exponent $\sim 0.6$.\\
\subsection{Proof of Thm. \ref{bds}}
\noindent
In the following $\mu>0$ is assumed, with no loss of generality. Lemma \ref{singmat} establishes a necessary condition in order
that an eigenvalue of ${\bf S}(\vartheta,\mu)$ for fixed $\mu>0$ may be constant for $\vartheta$ in a set of positive measure in $(0,2\pi/Q)$. Lemma \ref{nosuch} says that this condition cannot be met at any primitive resonance.
\begin{lemma}
\label{singmat}
{Let $\tau$ and $\beta$ be as in Prop.\ref{dado}, and let $d$ denote the integer part of
$\tfrac{Q+1}2$. If $\mu>0$ is fixed and the matrix ${\bf S}(\vartheta,\mu)$ has an eigenvalue which is constant for
$\vartheta$ in a subset of positive measure of $(0,\tfrac{2\pi}Q)$, then the matrix ${\bf G}^{(d)}:=\{G_{jk}\}_{1\leq j,k\leq d}$ is singular.}
\end{lemma}
{\it Proof:} eqn. (\ref{def1}) defines a matrix-valued function on the unit circle
$\rho=1$ in the plane of the complex variable $z=\rho e^{i\vartheta}$. This function is analytically continued to $\CM\setminus\{0\}$, in the form ${{\bf S}}(z)\;=\;{\bf Z}(z) {{\bf G}}$,
where $\bf G$ is as in (\ref{def2}), and ${\bf Z}(z)$ is a diagonal matrix, with diagonal elements  given by:
\begin{gather}
{Z}_{jj}(z)\;=\;
e^{-i\frac{\mu}2\left(z\xi^{1-j}+z^{-1}\xi^{j-1}\right)}\nonumber\\
=\;e^{-i\frac{\mu}2(\rho+\rho^{-1})
\cos\left(2\pi(j-1)/Q+\vartheta\right)}e^{\frac{\mu}2(\rho-\rho^{-1})\sin\left(2\pi(j-1)/Q+\vartheta\right)}
\;.
\label{ancont}
\end{gather}
where $\xi:=e^{-2\pi i/Q}$.
Whenever $\vartheta$ is not a multiple of $\pi/Q$, this matrix is hyperbolic.
In particular, for $z_{\rho}:=\rho e^{i\tfrac {\pi}{2Q}}$
and $\rho>1$  the expanding subspace of ${\bf Z}(z_{\rho})$ has dimension $d$ given by the integer part
of $\tfrac{Q+1}{2}$. Vectors $\bf x$ in the expanding subspace have
components $x_j=0$ for $d<j\leq Q$ and projection onto such vectors will be denoted $\bf P$.  \\
 Let $w$, ($|w|=1$),  be a constant eigenvalue of ${\bf S}(z)$ for all $z$ in a subset of positive measure of the circle $\{|z|=\rho=1\}$.
 For any fixed value $\lambda\in\CM$ the characteristic polynomial  ${\mathcal P}(z,\lambda):=\det({\bf S}(z)-\lambda)$ is an analytic function of $z$ in
$\CM\setminus\{0\}$. This is in particular true for $\lambda=w$, and then analyticity forces ${\mathcal P}(z,w)$ to vanish at all points $z$ in $\CM\setminus\{0\}$. Thus $\forall\rho>1$
$w$ is a unimodular eigenvalue of ${{\bf S}}(z_{\rho})$, and
there exists ${\bf x}_{\rho}\in\CM^Q$ with $||{\bf x}_{\rho}||=1$ such that
${{\bf S}}(z_{\rho}){\bf x}_{\rho}=w{\bf x}_{\rho}$.
Then
$||{{\bf P}}{{\bf S}}(z_{\rho}){\bf x}_{\rho}||=||{{\bf P}}{\bf x}_{\rho}||\leq 1
$
, whence, using (\ref{ancont}) :
\begin{gather}
\label{sad1}
1\;\geq\; ||{{\bf P}}{{\bf S}}(z_{\rho}){\bf x}_{\rho}||\;=\;||{\bf Z}(z_{\rho}){\bf P}{\bf G}{\bf x}_{\rho}||\;\geq\;
e^{c(\rho-\rho^{-1})}||{{\bf P}}{{\bf G}}{\bf x}_{\rho}||
\;,
\end{gather}
where $c:=\mu\sin(\pi/(2Q))$.
Similarly, from
$||{{\bf P}^{\bot}}{{\bf S}}(z_{\rho}){\bf x}_{\rho}||=||{{\bf P}^{\bot}}{\bf x}_{\rho}||$
it follows that
\begin{gather}
\label{sad2}
||{{\bf P}}^{\bot}{\bf x}_{\rho}||\;\leq\; e^{-c(\rho-\rho^{-1})}||{{\bf P}}^{\bot}{\bf G}{\bf x}_{\rho}||\;\leq
\;e^{-c(\rho-\rho^{-1})}
\;.
\end{gather}
As $\rho$ is arbitrarily large, (\ref{sad1}) and (\ref{sad2}) imply that there is a $\bf y$ of unit norm, such that ${\bf P}{\bf y}
={\bf y}$ and ${\bf P}{\bf G}{\bf y}=0$. As the range of $\bf P$
is the expanding subspace of dimension $d$ , the claim is proven. $\Box$.
\begin{lemma}
\label{nosuch}
{Let $\tau$, $\beta$ be as in Prop. \ref{dado}, and let $d$ and  ${\bf G}^{(d)}$ be as in Lemma \ref{singmat}. If $Q=q$, {\it i.e.} $P$ and $Q$ are coprime, then the matrix ${\bf G}^{(d)}$  is not singular.}
\end{lemma}
{\it Proof :}
For $q=1$ the claim is obvious, so $q>1$ will be assumed. For $1\leq j\leq q$ let ${\bf e}^{(j)}$ the vector in $\CM^{q}$ whose $r$-th component is $e^{(j)}_r=\xi^{(j-1)(r-1)}$, where $\xi:=e^{-2\pi i/q}\neq 1$. Let ${\bf y}=(y_1,\ldots,y_d)\in\CM^d$ satisfy  ${\bf G}^{(d)}{\bf y}=0$, and ${\bf y}\neq 0$. From Definition (\ref{def2}) it follows that the vector ${\bf b}\in\CM^{q}$ which is defined by:
$$
{b}_r\;=\;a_r\sum\limits_{k=1}^d y_k{e}^{(k)}_r\;,\;\;1\leq r\leq q
$$
is orthogonal to ${\bf e}^{(j)}$ for $1\leq j\leq d$. As vectors $\{{\bf e}^{(j)}\}_{1\leq j\leq q}$ are a basis in $\CM^{q}$,  $\bf b$ must be a combination of
vectors ${\bf e}^{(j)}$ with $d+1\leq j\leq q$; and so, constants $u_k$, $(1\leq k\leq q-d)$ exist, so that
\begin{equation}
\label{arr}
a_r\sum\limits_{k=1}^d y_k\;\xi^{(k-1)(r-1)}\;=\;\xi^{d(r-1)} \sum\limits_{k=1}^{q-d} u_k\xi^{(k-1)(r-1)}\;\;,
\;\;\;1\leq r\leq q\;.
\end{equation}
Defining polynomials  $F(z)=\sum_1^{d}y_kz^{k-1}$ and  $G(z)=\sum_{1}^{q-d}u_kz^{k-1}$, eqn. (\ref{arr}) may be rewritten as:
\begin{equation}
\label{arr1}
G(\xi^{r-1})\;=\;c_r F(\xi^{r-1})\;\;,\;\;c_r:=a_r\xi^{d(1-r)}\;\;,\;\;1\leq r\leq q\;.
\end{equation}
Replacing $r$ with $r+s$ in this equation  one obtains :
\begin{equation}
\label{arr11}
c_{r+s}\;G(\xi^{r-1})\;F(\xi^{r+s-1})\;=\;c_r\;G(\xi^{r+s-1})\;F(\xi^{r-1})\;.
\end{equation}
Then, using (\ref{def3}) and (\ref{arr1}):
\begin{equation}
\label{arr2}
\frac{c_{r+s}}{c_s}\;=\;e^{i\pi \tfrac{s}{q}(2d-ps-2p\beta)}\;e^{-2\pi is(r-1)\tfrac pq}\;,
\end{equation}
Since $p,q$ are mutually prime, $s$ may be chosen such that $ps=1\mod (q)$, and then:
$$
\frac{c_{r+s}}{c_r}\;=\;e^{i\gamma}\;\xi^{r-1}\;,
$$
where $\gamma$ is independent of $r$. Denoting $\lambda:=\xi^s$, from (\ref{arr11}) it follows that
\begin{equation}
\label{arr3}
e^{i\gamma}\;z \;G(z)\;F(\lambda z)\;-\;G(\lambda z)\; F(z)\;=\;0
\end{equation}
whenever $z$ is a $q$-th root of unity, {\it i.e.} $z=\xi^{r-1}$, $1\leq r\leq q$. By definition of $F(z)$ and $G(z)$,  the polynomial on the lhs in (\ref{arr3}) has degree at most $q-1$  and so it must identically vanish. Hence, $\forall z$:
\begin{equation}
\label{arr4}
e^{i\gamma}\;z \;G(z)\;F(\lambda z)\;=\;G(\lambda z)\; F(z)\;.
\end{equation}
As $G(z)F(\lambda z)$ and  $G(\lambda z) F(z)$ have the same degree, they must vanish in order that this equation be satisfied $\forall z$. This implies that either $F(z)=0$ or $G(z)=0$ identically. The latter case implies the former because then $F(z)$ has to vanish at the $q$-th roots of unity due to eqn. (\ref{arr1}), and its degree is $\leq d-1<q$. Hence
$F(z)$ vanishes identically in all cases, in contradiction to ${\bf y}\neq 0$. $\Box$


\subsection{Proof of Thm.\ref{bdwth} .}
\subsubsection{Perturbation Theory.}
\label{pth}

\noindent For the purposes of this section it is convenient to unitarily transform the matrix ${\bf S}(\theta,\mu)$ to ${\bf X}(\theta,\mu):= {\bf F}(\theta)^{-1}{\bf G}{\bf S}(\vartheta/q,\mu){\bf G}^{-1}{\bf F}
(\theta)$, where the unitary matrix ${\bf F}$ is defined by the matrix elements:
$$
{F}_{jk}(\theta)\;=\;\tfrac 1{\sqrt{q}}\;e^{-i(j-1)\theta/q}e^{-2\pi i(j-1)(k-1)}\;\;,\;\;1\leq j,k\leq q\;.
$$
Straightforward calculation shows that :
\begin{equation}
\label{defop}
 {\hat{\bf X}}(\vartheta,\mu)\;=\;{\hat{\bf
C}}\;e^{-i\mu{\hat{\bf V}}(\vartheta)}\;.
\end{equation}
where
\begin{gather}
{\hat{\bf C}}\;=\;\sum\limits_{j=1}^q e^{-\pi ip(
j-1/2)^2/q}\;|j\rangle\langle j|\;,\label{eigvsc}\\
{\hat{\bf V}}(\vartheta)\;=\;\frac12\sum\limits_{j=1}^{q-1}
\left(\;|j\rangle\langle j+1|\;+\;|j+1\rangle\langle j|\;\right)\;+
\;\frac12\left(\;|1\rangle\langle
q|e^{i\vartheta}+|q\rangle\langle 1|e^{-i\vartheta}\;\right)\;;\label{expv}
\end{gather}
Dirac notations are used, and $\{|j\rangle\;,\;j=1,2,\ldots,q\}$
denotes the canonical basis in $\CM^q$.

\noindent Let $\gamma$ be a circular
path in the complex plane, counterclockwise oriented, centered at
an unperturbed eigenvalue $a_j$, and containing no other
unperturbed eigenvalue. For sufficiently small $\mu<2\pi/q$, this circle will contain
no perturbed eigenvalue, except those which were born of $a_j$.  Then
\begin{equation}
\label{res} {\Pi}(\gamma,\theta,\mu)\;=\; -\frac1{2\pi
i}\int_{\gamma} dz\;({{\bf X}}(\theta,\mu)\;-z)^{-1}\;
\end{equation}
is projection onto the eigenspace of ${\bf X}$ which corresponds
to these eigenvalues.
Consequently,
\begin{equation}
\label{res1} { \Pi}(\gamma,\theta,\mu)\;{{\bf
X}}(\theta,\mu)\;{\Pi}(\gamma,\theta,\mu)\;=\;-
\frac1{2\pi i}\int_{\gamma}dz\;z\;({{\bf X}(\theta,\mu)}\;-\;z)^{-1}\;.
\end{equation}
We denote
$$
{{\bf R}}(z)\;=\;[{{\bf C}}-z{{\bf
I}}]^{-1}\;\;,\;{{\bf L}}(z)\;=\;{{\bf R}}(z){{\bf
C}}\;.
$$
For sufficiently small $\mu$, the resolvent in (\ref{res}) may be
expanded as follows:
\begin{equation}
\label{resop}
\begin{split}
({{\bf X}}(\theta,\mu)-z)^{-1}= \left[ {{\bf I}}+
{{\bf L}}(z)(\;e^{-i\mu{{\bf V}}(\theta)}-{{\bf
I}}\;)\;\right]^{-1}{{\bf R}}(z)\\
=\;{{\bf R}}(z)\;+\;\sum\limits_{n=1}^{\infty}(-1)^n \left[
{{\bf L}}(z)(\;e^{-i\mu{{\bf V}}(\theta)}-{{\bf
I}}\;)\; \right]^n{{\bf R}}(z)
\end{split}
\end{equation}
Next we expand the exponential in powers of $\mu$. Let
$\Omega_{n,{\ell}}$ denote the set of vectors
$\overrightarrow{r}^{(n)}\in\NM^n$ whose components $r_1,\ldots,r_n$, are
strictly positive integers that satisfy
$|\overrightarrow{r}^{(n)}|=r_1+\ldots+r_n=\ell$. For
$\overrightarrow{r}^{(n)}\in \Omega_{n,\ell}$ let
\begin{equation}
\label{pprod} {{\bf P}}(\theta,z,\overrightarrow{r}^{(n)})\;=\;
{{\bf L}}(z) {{\bf V}}(\theta)^{r_1}\;{{\bf
L}}(z){{\bf V}} (\theta)^{r_2}\;\ldots\;{{\bf
L}}(z){{\bf V}}(\theta)^{r_n}{{\bf L}}(z){{\bf
C}}^{-1}\;.
\end{equation}
With such notations, one may expand the resolvent in eqn.
(\ref{res}) in powers of $\mu$ as follows:
\begin{equation}
\label{expans}
 ({{\bf X}}(\theta,\mu)-z)^{-1}=\;
 {{\bf R}}(z)\;+\;\sum\limits_{\ell=1}
^{\infty}(-i)^{\ell}\mu^{\ell}\;{{\bf Q}}_{\ell}(\theta,z)\;,
\end{equation}
where:
\begin{equation}
\label{expli} {{\bf Q}}_{\ell}(\theta,z)\;=\;
\sum\limits_{n=1}^{\ell}(-1)^n\sum\limits_{\overrightarrow{r}^{(n)}
\in\Omega_{n,\ell}} \frac 1{r_1! r_2!\ldots r_n!}{{\bf
P}}(\theta,z,\overrightarrow{r}^{(n)})\;\;.
\end{equation}
 Replacing (\ref{expans}) in (\ref{res}) yields  the
expansion of the spectral projector (\ref{res}) in powers of $\mu$.
Matrix elements of the operator ${\bf P}$ between arbitrary
states $|j\rangle$ and $|k\rangle$ may be  written as sums over
paths, specified by strings ${\pmb m} \equiv(m_0,\ldots,m_{\ell})$
of integers taken from $\{1,2,\ldots,q\}$, such that $m_0=j$ and
$m_{\ell}=k$. Moreover, from the explicit form of ${\bf V}$ it
is apparent that only those paths contribute, which jump by $\pm
1$(mod $q$) at each step. The set of such paths will be denoted by
$\Lambda_{}(\ell,j,k)$. For ${\pmb m}\in\Lambda_{}(\ell,j,k)$, let
the integer $\nu({\pmb m})$ count the number of jumps $q\to 1$,
minus the number of jumps $1\to q$. Then, using (\ref{expv}):
\begin{equation}
\label{p1q}
\begin{split}
\langle j|{{\bf P}}(\theta,z,\overrightarrow{r}^{(n)})|k\rangle=
\sum\limits_{{\pmb m}\in\Lambda(\ell,j,k)}
G({\pmb m},\overrightarrow{r}^{(n)},z)\;\times\\
\times\langle m_0|{{\bf V}}(\theta)|m_1\rangle\langle
m_1|{{\bf V}}(\theta)|m_2\rangle\ldots\langle
m_{\ell-1}|{{\bf V}}(\theta) |m_{\ell}\rangle\;
\\
=\;2^{-\ell}\sum\limits_{{\pmb m}\in\Lambda(\ell,j,k)} G({\pmb
m},\overrightarrow{r}^{(n)},z)\;e^{i\nu({\pmb m})\theta}\;,
\end{split}
\end{equation}
where
\begin{equation}
\begin{split}
\label{gigi} G({\pmb m},\overrightarrow{r}^{(n)},z)\;=\;g(m_{0},z)\;
g(m_{r_1},z)g(m_{r_1+r_2},z)\ldots g(m_{\ell},z)h(m_{\ell})\;,\noindent\\
g(m,z )=a_m/(a_m-z)\;\;,\;\;h(m)=a_m^{-1}\;.
\end{split}
\end{equation}
 From (\ref{p1q}) the definitions of
${\bf L}$, ${\bf V}$, and ${\bf C}$ it follows that :
\begin{gather}
\label{reverse} \langle j|\;{{\bf
P}}(\theta,z,\overrightarrow{r}^{(n)})\;|k\rangle\;=\;
\langle k|\;{{\bf
P}}(2\pi-\theta,z,\overrightarrow{r}'^{(n)})\;|j\rangle\
\end{gather}
where $\overrightarrow{r}'^{(n)}$ is  the reverse of $\overrightarrow{r}^{(n)}$,
that is, $r'_j=r_{n+1-j}$, $(1\leq j\leq n)$.
\subsubsection{Degenerate case. }
\noindent Let
$j< (q+1)/2$, so that $a_j=a_{j'}$ ($j'=q-j+1$).  Let
$$
|j_+\rangle\;=\;\frac1{\sqrt 2}\,{ \Pi}\;(\gamma,\theta,\mu)
\left(|j\rangle\;+\;|j'\rangle\right)\;\;\;,\;\;\;
|j_-\rangle\;=\;\frac1{\sqrt 2}\,{ \Pi}(\gamma,\theta,\mu)\left(
|j\rangle\;-\;|j'\rangle\right);\;\;.
$$
Gram-Schmidt orthonormalization on $|j_{\pm}\rangle$ yields
orthonormal vectors $|{\tilde j}_{\pm}\rangle$ and we shall
calculate the corresponding $2\times 2$ matrix of ${{\bf
X}}(\theta,\mu)$. To this end we shall compute matrix element of the operators ${
\Pi}(\gamma,\theta,\mu)$ and ${
\Pi}(\gamma,\theta,\mu){{\bf X}}{
\Pi}(\gamma,\theta,\mu)$ using eqs.(\ref{res}) and (\ref{res1}), along with expansion (\ref{expans}),
(\ref{expli}),(\ref{p1q}),(\ref{gigi}). Such elements may be written as the sum of a
$\theta$-independent part (given by the average over
$\theta$), plus a $\theta$-dependent part. The former part is determined
by paths ${\pmb m}$ with index $\nu({\pmb m})= 0$. The
shortest such path from $j$ to $j'$ is :
\begin{equation} \label{shortright} \overrightarrow{{\pmb
m}_j}:=(j,j+1,\ldots,j'-1,j')\;,\;\;\;\mbox{\rm  of length}\;\;
q-2j+1.
\end{equation}
  The latter part is determined
by paths ${\pmb m}$ with index $\nu({\pmb m})\neq 0$, and the
shortest such path from $j$ to $j'$ is
\begin{equation}
\label{shortleft} {\overleftarrow{\pmb m}_j}:=(j,j-1,\ldots,1,
q, q-1,\ldots, j') ,\;\;\;\mbox{\rm of length}\;\;2j-1 .
\end{equation}
with $\nu(\overleftarrow{{\pmb m}_j})=1$. On the other hand, the shortest loops with $\nu\neq 0$ from $j$ to
$j$ (or from $j'$ to $j'$) have length $q$. There are 2 such loops, reverse of each other .
Hence we may write:
\begin{gather}
\label{s1}\langle j|{ \Pi}(\gamma,\theta,\mu)|j\rangle\sim
1+D_j(\mu)+\mu^q c_j\cos(\theta)+o(\mu^q)\;,\\
\label{s12}\langle j|{
\Pi}(\gamma,\theta,\mu)|j'\rangle\sim F_j(\mu)+\mu^{2j-1}
u_je^{i\theta}+o(\mu^{2j-1})\;,
\\\label{s3} \langle j|{
\Pi}(\gamma,\theta,\mu){{\bf X}}{
\Pi}(\gamma,\theta,\mu)|j\rangle\sim a_j +
G_j(\mu)+\mu^q b_j\cos(\theta)+o(\mu^q)\;,\\
\label{s4}\langle j|{
\Pi}(\gamma,\theta,\mu){{\bf X}}{
\Pi}(\gamma,\theta,\mu)|j'\rangle\sim
H_j(\mu)+\mu^{2j-1}v_je^{i\theta}+o(\mu^{2j-1})\;,
\end{gather}
where $D_j(\mu), F_j(\mu), G_j(\mu)$ and $H_j(\mu)$ are independent
of $\theta$, and what follows on their right is the
$\theta$-dependent parts. To leading orders,
\begin{equation}
\label{s2} D_j(\mu)\;\sim\;d_j\;\mu^2\;\;,\;\; F_j\;\sim\;
f_j\;\mu^{q-2j+1}\;\;,\;\; G_j\;\sim\;g_j\;\mu^2\;\;,\;\;
H_j\;\sim\;h_j\;\mu^{q-2j+1}\;.
\end{equation}
Coefficients $f_j$ and $h_j$
are computed by picking the contribution of path
(\ref{shortright}) from  eqn. (\ref{p1q}), by inserting it   in eq.
(\ref{expli}) and then in eq. (\ref{expans}), and finally
integrating
along $\gamma$.   Coefficients $u_j$ and $v_j$ are computed
in a similar way, using path (\ref{shortleft}); and coefficients $d_j$ and $g_j$
are determined by loops of length $2$.
 Using (\ref{s1}) and (\ref{s2}) we find:
\begin{equation}
\begin{split}
\label{matr1} \langle j_{\pm}|{
\Pi}(\gamma,\theta,\mu)|j_{\pm}\rangle\;\sim\;
1\;+\;D_j(\mu)\;\pm\; F_j(\mu)\;\pm\;u_j\mu^{2j-1}\;\cos(\theta)\;,\\
\langle j_{\pm}|{
\Pi}(\gamma,\theta,\mu)|j_{\mp}\rangle\;\sim\;
\mp\;iu_j\mu^{2j-1}\;\sin(\theta)\;,
\end{split}
\end{equation}
and, similarly,
\begin{gather}
\label{matr2} \langle j_{\pm}|{
\Pi}(\gamma,\theta,\mu){{\bf X}}{
\Pi}(\gamma,\theta,\mu)|j_{\pm}\rangle\;\sim\;
a_j\;+\;G_j(\mu)\;\pm\;H_j(\mu)\;\pm\;\mu^{2j-1}\;v_j\cos(\theta)\;,\\
\langle j_{\pm}|{ \Pi}(\gamma,\theta,\mu){{\bf X}}{
\Pi}(\gamma,\theta,\mu)|j_{\mp}\rangle\;\sim\;
\mp\;i\mu^{2j-1}\;v_j\;\sin(\theta)\;.
\end{gather}
Using  all the above formulae, one computes
\begin{gather}
\langle{\tilde j}_{\pm}|{{\bf X}}|{\tilde
j}_{\pm}\rangle\;\sim\;
K_{j\pm}(\mu)\;\pm\; \mu^{2j-1} t_j\;\cos(\theta)\;,\nonumber\\
\langle{\tilde j}_{\pm}|{{\bf X}}|{\tilde
j}_{\mp}\rangle\;\sim\;\mp i\mu^{2j-1}t_j\;\sin(\theta)\;,
\label{matr3} \end{gather}
where:
\begin{gather}
 K_{j\pm}(\mu)\;=\; (a_j\;+\;G_j(\mu)\;\pm\;H_j(\mu))(
1\;-\;D_j(\mu)\;\mp\;F_j(\mu))\;,\nonumber\\
t_j\;=\;v_j\;-\;a_j\;u_j\;.\label{matr4}
\end{gather}
Let $\xi^{\pm}_j=\xi_j^{\pm}(\theta,\mu)$ denote the
eigenvalues of the matrix in (\ref{matr3}), labeled so that $w_j(\theta,\mu)
\sim\xi_j^{+}(\theta,\mu)$ and $w_{j'}(\theta,\mu)
\sim\xi_j^{-}(\theta,\mu)$ as $\mu\to 0$.
\begin{equation}
\label{eig} \frac{d\xi^{\pm}_j}{d\theta}\;\sim\; \mp\frac
{\mu^{2j-1}t_j\Delta_j(\mu)}{
\xi^{+}_j\;-\;\xi^{-}_j}\;\sin(\theta)\;,
\end{equation}
where, in the leading order,
\begin{equation}
\label{del}
\Delta_j(\mu)\;=\;K_{j+}(\mu)-K_{j-}(\mu)\;\sim\;
2(h_j-a_jf_j)\;\mu^{q-2j+1}\;.
\end{equation}
as follows from (\ref{s2}) and (\ref{matr4}). Moreover,
\begin{gather}
\label{eigh} \xi^{+}_j\;-\;\xi^{-}_j\;=\;\sqrt{
\Delta_j(\mu)^2\;+\;4\mu^{2j-1} t_j \Delta_j(\mu)
\cos(\theta)\;+\;4\mu^{4j-2}t_j^2}\;.
\end{gather}
On account of (\ref{del}), the leading term
in (\ref{eigh}) is $\Delta_j(\mu)$ whenever $2j-1>q-2j+1$,
and is instead $2t_j\mu^{2j-1}$, in the opposite case.
Replacing in eqn.(\ref{eig}) we  find that
\begin{equation}
\label{fin1} \frac{d\xi^{\pm}_j}{d\theta}\;\sim\;\mp
\mu^{\alpha(j)}s_j\;\sin(\theta)\;,
\end{equation}
where $\alpha_j$ is as in (\ref{claim0}), and
\begin{gather}
s_j= \left\{
                                        \begin{array}{ll}
v_j-a_ju_j,
& \hbox{if $2j-1>q-2j+1$;} \\
                                          h_j-a_jf_j ,
& \hbox{if $2j-1<q-2j+1$.}
                                        \end{array}
                                      \right.
                                      \label{sj}
\end{gather}

\subsubsection{Nondegenerate case .}
\noindent In the case  when $j=(q+1)/2$,   $a_j$
is a nondegenerate eigenvalue for $\mu=0$, and then Eqs. (\ref{s1}) and (\ref{s3}) lead to:
\begin{gather}
\label{ndeg}
w_j(\theta,\mu)\;=\;\frac{
\langle j|{
\Pi}(\gamma,\theta,\mu){{\bf X}}{
\Pi}(\gamma,\theta,\mu)|j\rangle\
}
{
\langle j|{ \Pi}(\gamma,\theta,\mu)|j\rangle\
}
\sim\;a_j+G_j(\mu)-a_jD_j(\mu)+\mu^q s_j\cos(\theta)\;,
\end{gather}
where $s_j=b_j-a_jc_j$.

\subsubsection{Proof of Estimate (\ref{claim3})}
\noindent From  eqs.(\ref{p1q}),(\ref{gigi}), (\ref{s1}), and  (\ref{sj})  it
follows that:
\begin{equation}
\label{est1} s_j\;=\; \frac{1}{2\pi }\frac1{2^{\alpha_j}}
\sum\limits_{n=1}^{\alpha_j} (-1)^{n+j}
\sum\limits_{\overrightarrow{r}^{(n)}
\in\Omega(n,\alpha_j)}\frac1{\overrightarrow{r}^{(n)}!} \int_{\gamma}dz\;
(z-a_j)\;G({\pmb m}^{\uparrow}_j,\overrightarrow{r}^{(n)},z)\;,
\end{equation}
where $\overrightarrow{r}^{(n)}!$ is shorthand for $r_1!\ldots r_n!$,
$\alpha_j$ is given in eqn.(\ref{claim0}), and  ${\pmb m}^{\uparrow}_j$ is either the
path $\overrightarrow{{\pmb m}_j}$ (\ref{shortright})
(when
$j<(q+2)/4$ ), or the path $\overleftarrow{{\pmb m}_j}$ (\ref{shortleft}) (when $j\geq(q+2)/4$). The following proof is for the former path and is split in Lemmata \ref{l1} and \ref{l2} below.  The proof for the latter path is essentially identical, apart from notations.
 \begin{lemma}
\label{l1}
{: There are constants $C>0$, $\gamma >0$ so that
\begin{equation}\label{lm}
|s_j|\;\lesssim\; Cq^{1/2}\;e^{-\gamma q}\prod\limits_{l=1}^{\alpha_j-1}\frac1{|a_{j+l}a_j^{*}-1|}\;.
\end{equation}
}
\end{lemma}
{\it Proof:}
using definition (\ref{gigi}), and Cauchy's Integral formula,
\begin{equation}
\label{est2}
 \frac1{2\pi i}\int_{\gamma}dz\;
(z-a_j)\;G(\overrightarrow{m}_j,\overrightarrow{r}^{(n)},z)\;=\;
-\frac{a_j a_{j+l_1}\ldots a_{j+l_{n-1}}}
{(a_{j+l_1}-a_j)\ldots(a_{j+l_{n-1}}-a_j)}\;,
\end{equation}
where $l_s=r_1+\ldots+r_s$. Replacing this in (\ref{est1}),
\begin{equation}
\label{est3}
|s_j|\;\leq\;\frac1{2^{\alpha_j}}\sum\limits_{n=1}^{\alpha_j}\Xi_n\;,
\end{equation}
where:
\begin{equation}
\label{est4}
\Xi_n:=\sum\limits_{\overrightarrow{r}^{(n)}
\in\Omega(n,\alpha_j)}
{\overrightarrow{\mathcal G}}_n(j,\overrightarrow{r}^{(n)})\;\;\;\mbox{\rm and}\;\;\;
{\mathcal G}_n(j,\overrightarrow{r}^{(n)}):=
\frac1{{\overrightarrow{r}^{(n)}}!}\prod\limits_{s=1}^{n-1}\frac{1}{|a_{j+l_s}a_j^{*}-1|}\;,
\end{equation}
If  $\overrightarrow{r}^{(n)}
\in\Omega(n,\alpha_j)$ and $n<\alpha_j$, then there is $s'\in\{1,\dots,n\}$ such that
$r_{s'}$ is the sum of two positive integers $t',t''$.
Define $\overrightarrow{r}^{(n+1)}:=(r_1,\ldots,r_{s'-1},t',t'',r_{s'+1},\ldots, r_n)\in\Omega(n+1,\alpha_j)$, and $\omega^{(n+1)}=a_{j+l_{s'-1}+t'}a_j^{*}$.  Then
\begin{gather}
{\mathcal G}_{n+1}(j,\overrightarrow{r}^{(n+1)})=
{\mathcal G}_n(j,\overrightarrow{r}^{(n)})
 \binom{r_{s'}}{t'}
 \frac1{|\omega^{(n+1)}-1|}\;\geq\;\frac2{|\omega^{(n+1)}-1|}
{\mathcal G}_n(j,\overrightarrow{r}^{(n)})\;,
\end{gather}
Repeating the argument, a sequence
$\omega^{(n+1)},\ldots,\omega^{(\alpha_j-1)},\omega^{(\alpha_j)}$ can be constructed, so that
\begin{equation}\label{est5}
{\mathcal G}_n(j,\overrightarrow{r}^{(n)})\leq 2^{n-\alpha_j}{\mathcal G}_{\alpha_j}(j,\overrightarrow{r}^{(\alpha_j)})
\prod\limits_{l=n+1}^{\alpha_j}|\omega^{(l)}-1|\;,
\end{equation}
where $\overrightarrow{r}^{(\alpha_j)}=(1,1,\ldots,1)$ is the one
vector in $\Omega(\alpha_j,\alpha_j)$.
Denote ${\mathcal P}(j,n,\rn)$ the product on the rhs
in (\ref{est5}).
All numbers $\omega^{(l)}$ are $q$-th roots of
unity, so  ${\mathcal P}(j,n,\rn)$ is a product of diagonals in the regular
$q$-gon inscribed in the unit circle in the complex plane, drawn
from the vertex at $1$. The largest possible value of
such a product is attained, if the vertices $\neq 1$ of
the diagonals are taken as close as possible to $-1$; keeping in
mind, that each diagonal may occur  twice, and at most twice,  in the
product, due to the symmetry $a_j=a_{q-j+1}$; and that two different
diagonals may have the same length, because
for odd $q$ the regular $q$-gon is symmetric with respect to any
diameter drawn through one of the vertices. It follows that
\begin{equation}
\label{est6}
{\mathcal P}(j,n,\rn)\;<\;2^{\alpha_j-n}\prod\limits_{m=1}^{L(j,n)}\cos^4 ((m-1/2)\pi/q)\;,
\end{equation}
where $L(j,n):=$ Int$((\alpha_j-n)/4)$; therefore,
\begin{gather}
\label{est7}
{\mathcal P}(j,n,\rn)\;\lesssim\;2^{\alpha_j-n}C_1e^{qF(4L(j,n)/q)}\;,
\end{gather}
where $C_1>0$ is a numerical constant, and
\begin{equation}
\label{est71}
F(x)\;=\;\int_0^x dt\;\log(\cos(\pi t/4))\;.
\end{equation}
From (\ref{est3}), (\ref{est4}), (\ref{est5}), and (\ref{est7}) :
\begin{equation}
\label{estt}
|s_j|\;<\;C_1{\mathcal G}_{\alpha_j}(j,\overrightarrow{r}^{(\alpha_j)})\sum\limits_{n=1}^{\alpha_j}
\frac1{2^{\alpha_j}}\binom{\alpha_j-1}{n-1}
e^{qF(4L(j,n)/q)}
\;,
\end{equation}
because there are
$\bigl(\begin{smallmatrix}\alpha_j-1\\n-1\end{smallmatrix}\bigr)$ elements in $\Omega(n,\alpha_j)$.
The sum on the rhs in (\ref{estt}) is an average over a binomial distribution. Since $\alpha_j>q/2$ (see eqn.(\ref{claim0})), as $q\to\infty$ this
distribution approaches a Gaussian distribution with mean $\alpha_j/2$ and variance $\alpha_j/4$, so, replacing   $(\alpha_j-n)/q$ by a continuous variable $x$, and using again that $q>\alpha_j>q/2$,
\begin{equation}
\label{est8}
|s_j|\;\lesssim\;C_2q^{1/2}{\mathcal G}_{\alpha_j}(j,\overrightarrow{r}^{(\alpha_j)})\int_0^1dx\;e^{q[-2(x-\lambda_j)^2+F(x)]}\;,
\end{equation}
where $\lambda_j=\alpha_j/(2q)$, and so $1/4<\lambda_j<1/2$. Analysis  of the function
$-2(x-\lambda)^2+F(x)$ in $(x,\lambda)\in[0,1]\times(1/4,1/2)$ yields the upper bound $-\gamma\approx-0.0016$.
This proves the Lemma, because
${\mathcal G}_{\alpha_j}(j,\overrightarrow{r}^{(\alpha_j)})$ in (\ref{est8}) is just the product which appears
in (\ref{lm}).
$\Box$\\
We are left with estimating the product on the rhs of (\ref{lm}):
\begin{lemma}
\label{l2}
{For $q\to\infty$,
$$
\left\vert\sum\limits_{l=1}^{\alpha_j-1}
\log\left(|a_{j+l}a_j^*-1|\right)\right\vert\;=\;O\left(\sqrt{q\log^3(q)}\right)\;.
$$}
\end{lemma}
{\it Proof}: define:
\begin{equation}
\label{rho}
{\mathcal S}(\phi,\rho)\;:=\;\log(|1-\rho e^{i\phi}|)\;=\;\sum\limits_{N\in\ZM}
\sigma_N(\rho)\;e^{iN\phi}\;\;\;,\;\;(0<\rho\leq 1\;,\;0\leq \phi\leq 2\pi)\;,
\end{equation}
where \cite{GR}:
\begin{equation}
\label{cdf}
\sigma_N(\rho)\;=\;-\frac{\rho^{|N|}}{|N|}\;\;\;(N\neq 0)\;\;\;\;,\;\;\sigma_0(\rho)\;=\;0
\end{equation}
From eq.(\ref{rho}):
\begin{equation}
\log\left(\prod\limits_{l=1}^{\alpha_j-1}|\rho a_{j+l}a_j^{*}-1|\right)\;=\;
\sum\limits_{N\in\ZM}\sigma_N(\rho)\sum\limits_{l=1}^{\alpha_j}a_{j+l}^Na_j^{N*}\;.
\end{equation}
We write the sum over $N$ as  $\Sigma'+\Sigma''$, where $\Sigma'$ is the sum restricted to $N\in q\ZM$.
In order to estimate $\Sigma'$ and $\Sigma''$ standard facts about sums of the Gauss
type are used. These are reviewed in Lemma \ref{l3} below. Using that $\alpha_j\leq q$, and
$\sigma_N<0$ for $N\neq 0$,
\begin{gather}
|\Sigma''|\;=\;|\sideset{}{''}\sum\sigma_N(\rho)\sum\limits_{l=1}^{\alpha_j-1}a_{j+l}^Na_j^{N*}|\;\leq\;
-\sqrt{2q\log(q)}\sideset{}{''}\sum\sigma_N(\rho)\nonumber\\
\leq\;
-\sqrt{2q\log(q)}\;{\mathcal S}(0,\rho)\;=\;
-\log(1-\rho)\;\sqrt{2q\log(q)})\;.
\label{sigma1}
\end{gather}
Using Lemma 2, and (\ref{cdf}), the remaining  sum is estimated as:
\begin{gather}
\label{sigma2}
\left\vert\Sigma'\right\vert\;\leq\;-q\sum\limits_{n\in\ZM}\sigma_{nq}(\rho)\;=\;
-2\log(1-\rho^q)\;.
\end{gather}
Finally, for $z$ complex with $|z|=1$, and $0<\rho<1$,
$$
\left\vert\log(|1-\rho z|)-\log(|1-z|)\right\vert\;\leq\;\frac{1-\rho}{|1-z|}
\;,
$$
whence it follows that:
\begin{gather}
\left\vert\log\left(\prod\limits_{l=1}^{\alpha_j-1}|a_{j+l}a_j^{*}-1|\right)-\log\left
(\prod\limits_{l=1}^{\alpha_j-1}|\rho a_{j+l}a_j^{*}-1|\right)\right\vert\;\leq\;
\nonumber\\
\leq\;(1-\rho)\sum\limits_{l=1}^{\alpha_j-1}\frac1{|a_{j+l}a_j^{*}-1|}\;<\;
16q(1-\rho)\sum\limits_{r=1}^{q}r^{-1}\;<\;16q(1-\rho)\log(2q)\;.
\end{gather}
and so
$$
\left\vert\log\left(\prod\limits_{l=1}^{\alpha_j-1}|a_{j+l}a_j^{*}-1|\right)\right\vert\;\leq\;
\Sigma'+\Sigma''+16q(1-\rho)\log(2q)\;;
$$
The proof is concluded on substituting (\ref{sigma1}) and (\ref{sigma2}) in the last estimate, and choosing $1-\rho
=1/(q\log(q))$ .$\Box$
\begin{lemma}
\label{l3}
{Let $q$ prime, $p$ not a multiple of $q$, $N\in\ZM$, and $T$ an integer $\leq q$. Then:
$$
\left\vert\sum\limits_{n=j}^{j+T-1} a_n^N\right\vert^2\;\;\left\{
                                               \begin{array}{ll}
                                                 =\;T^2, & \hbox{\rm\small if $N$ is a multiple of $q$;} \\
                                                 =\;q,   & \hbox{\rm\small  if $N$ is not a multiple of $q$, and $T=q$;}\\
                                                 \leq\; 2q(1+\log(q)), & \hbox{\rm\small otherwise.}
                                               \end{array}
                                             \right.
$$}
\end{lemma}
 The 1st claim is obvious and the 2nd is well known.  If $N$ is not a multiple of $q$, and $T<q$, then:
\begin{gather}
\left\vert\sum\limits_{n=j}^{j+T-1}a_n^N\right\vert^2\;=\;
\left\vert\sum_{r=0}^{T-1}\sum_{s=0}^{T-1}e^{-i\pi pN(r-s)/q}\;e^{i\pi pN(r-s)(r+s)/q}\right\vert\\
=T\;+\;2\Re\sum_{h=1}^{T-1}e^{i\pi pN[h^2-h]/q}\;\sum_{s=0}^{T-1-h}e^{2i\pi psNh/q}\\
\leq T\;+\;4\sum\limits_{h=1}^{T-1}\frac{1}{|\sin((\pi pNh/q)|}\;\leq\;T\;+\;2\sum\limits_{h=1}^{T-1}\frac{q}
{(pNh)\mod(q)}\;.\end{gather}
As $pN$ is prime to $q$,  $pNh$ mod$(q)$ takes all values in $\{1,\ldots,q-1\}$
as $h$ varies from $1$ to $q-1$, and the claim follows from $T< q$.
$\Box$\\

\end{document}